\documentclass[a4paper,11pt]{article}
\pdfoutput=1 % if your are submitting a pdflatex (i.e. if you have
\usepackage{jcappub} % for details on the use of the package, please
\usepackage[T1]{fontenc} % if needed
\usepackage{adjustbox}

\title{\boldmath   GW190521 formation scenarios via relativistic  accretion}
\author[a,1]{Alejandro Cruz-Osorio,\note{Corresponding author.}}
\author[b,2]{Fabio D. Lora-Clavijo}
\author[c,3]{and Carlos Herdeiro}

\affiliation[a]{Institut f{\"u}r Theoretische Physik, Goethe Universit\"at, Max-von-Laue-Stra{\ss}e 1, 60438 Frankfurt, Germany}
\affiliation[b]{Grupo de Investigaci\'on en Relatividad y Gravitaci\'on, Escuela de F\'isica, Universidad Industrial de Santander A. A. 678, Bucaramanga 680002, Colombia}
\affiliation[c]{Departamento de Matem\'atica da Universidade de Aveiro and CIDMA, Campus de Santiago, 3810-183 Aveiro, Portugal}

\emailAdd{osorio@itp.uni-frankfurt.de}
\emailAdd{fadulora@uis.edu.co}
\emailAdd{herdeiro@ua.pt}

\abstract{The recent gravitational wave transient GW190521 has been interpreted by the LIGO-Virgo collaboration (LVC) as sourced by a binary black hole (BH) merger. According to the LVC parameter estimation, at least one of these progenitors falls into the so-called pair-instability  supernova mass gap. This raises the important question of how and when these progenitors \textit{formed}. 
In this paper we use an accretion model with super-Eddington mass accretion rate obtained from General Relativity hydrodynamics simulations to analyse the scenario wherein the GW190521 original progenitors (OPs) formed at lower masses (and spins) and grew to their estimated LVC parameters by relativistic accretion. We consider that the environment wherein the binary is immersed has density gradients as well as a dependence on the Mach number of the gas. Taking the LVC parameter estimation at $z=0.82$ as the endpoint of the accretion evolution, we  estimate the initial masses and spins of the OPs at three different red-shifts  $z=100, \ 50$, and $20$. We found three distinct possible types of OPs: $(i)$ $10^{-4} M_{\odot} - 3 M_{\odot}$ almost non-rotating (with Kerr spin parameter $a_{\star}< 10^{-2}$) primordial BHs;  $(ii)$ $3 M_{\odot} - 40M_{\odot}$ slowly rotating ($ 10^{-2} < a_{\star} < 0.5$) stellar mass BHs; $(iii)$ $40M_{\odot} - 70M_{\odot}$ BHs with a moderate spin parameter $a_{\star}\sim 0.5$, which could originate from the collapse of high mass Pop III stars. The mass spread is due to varying the density gradient and the relativistic Mach number of the cosmic plasma; the variation of the masses  due to the origin at different red-shifts, on the other hand, is negligible, $\sim 2\%$. 
For high Mach number scenarios, the BHs have low mass and spin accretion rates, leading to  OPs with masses and spins close to the GW190521 LVC estimated values.
We have also compared our results with previous studies where the Newtonian accretion model was used, finding relativistic corrections of   $\sim 13\%$ for the OPs masses. In particular, the relativistic model leads to smaller initial masses. 
} 

\begin{document}
\maketitle
\flushbottom

\section{Introduction}
\label{sec:intro}

The LIGO-Virgo scientific collaboration (LVC) has reported the detection of gravitational wave transients since 2016~\cite{Abbot2016-GW-detection-prl,Abbot2016g,Abbott2016c,Abbott2017f,Abbott2017g,Abbott2017_etal,Abbott2020,Abbott2020b}, opening up a new window towards highly relativistic  systems, such as neutron star (NS) and black hole (BH) binaries, and providing new information about their populations in the Cosmos. The mass of the objects inferred from these events spans a considerable range; for the final  object, for instance, the LVC reported  $2.74^{+0.04}_{-0.01} M_\odot$, for GW170817, interpreted as a NS-NS binary coalescence and $62^{+0.04}_{-0.01}M_\odot$ for GW150914, interpreted as a BH-BH binary merger. Yet,  the  masses of the progenitors in all events reported (until recently) correspond to the expected masses for NSs  or  stellar  mass BHs (sBHs). Recently, the LVC reported the transient GW190521~\cite{Abbott:2020tfl}, interpreted as the merger of two BHs with masses $85^{+21}_{-14} M_\odot$ and $66^{+17}_{-18} M_\odot$, which implies at least one  of these progenitors is  in the  so called pair-instability supernova mass gap \cite{Abbott:2020mjq}. This led to a debate on how the GW190521 progenitor BHs formed, and even on  the nature of the event, see $e.g.$ \cite{CalderonBustillo:2020srq, Sakstein:2020axg}. 

A natural scenario is that the progenitors of GW190521 started off as lower mass BHs and grew up to the LVC reported masses by  \textit{accretion}. A model to tackle gas dynamics around a point mass was developed by Bondi, Hoyle and Lyttleton (BHL) \cite{Hoyle1939,Bondi1944,Bondi52}. This model assumes accretion onto a central compact object moving in a homogeneously distributed gas and has been widely used in different relativistic astrophysical scenarios, where the accreting object corresponds to a BH or a NS. Several numerical works have been dedicated to studying the morphological patterns in the vicinity of the BH as well as the mass accretion rates  \cite{Petrich89,Font98a,Font98c,Font98d,Font1999b}.  Additionally, many numerical simulations of the BHL mechanism have been developed in the context of relativistic astrophysics  \cite{Donmez2010,Cruz2012,Lora2013,Cruz2013}, which have been made evermore realistic (and complex) by including magnetic fields \cite{Penner2011}, radiative terms \cite{Zanotti2011}, density and velocity gradients \cite{Lora2015219,Cruz2016}, small rigid bodies around the BH \cite{Cruz2017} and, more recently, by analyzing the common envelope phase in the evolution of binary systems \cite{cruz2020}, where useful relativistic expressions for the mass and angular momentum accretion rates, as well as luminosity, were obtained in terms of the density gradient parameter and Mach number. These expressions can be used to understand the behavior of the progenitors of the GW190521 event, at early stages, and so establish constraints on the initial masses and luminosity of this binary system in the relativistic regime.

Other recent works have also attempted to understand the GW190521 progenitors at early stages. For instance,  Ref. \cite{DeLuca2020a} argues that GW190521 cannot have been sourced by primordial BHs (PBHs) if PBHs do not accrete during their cosmological evolution. On the other hand, Ref. \cite{safarzadeh2020formation} proposes that the GW190521 binary system formation is possible due to gas accretion onto BH remnants from  Population III (Pop III) stars, born in high red-shift mini-halos. Moreover in  Ref. \cite{natarajan2020new} a gas accretion driven mechanism that can build up BH masses rapidly in dense, gas-rich nuclear star clusters is presented. In Ref. \cite{rice2020growth}, the authors show how the growth of sBH, embedded in a dense molecular cloud, can be associated with the GW190521 event. It is worth mentioning that in all these works the gas around the BH is described by the \textit{Newtonian} mass accretion formula. Another scenario is presented in Ref. \cite{palmese2020gw190521}, where the GW190521 progenitors are proposed to  be the result of the merger of the central BHs from two ultra–dwarf galaxies. Furthermore, the possibility that the GW190521 binary system comes from the first generation of massive, metal-free, Pop III stars, is studied in Ref.  \cite{liu2020population}.

Another interesting  aspect of  GW190521 is that the total mass of the GW190521 remnant is $142^{+28}_{-16} M_\odot$, which can be considered as an intermediate mass BH (IMBH).
IMBHs are an elusive class of BHs, albeit the logical intermediary between sBH and supermassive BHs (SMBH), with a  mass range from $10^2 M_\odot$ to $10^5 M_\odot$. If IMBHs are produced by accretion onto sBHs or mergers of massive stars or sBHs, then IMBHs should be abundant. On the other hand, if IMBHs are only produced from core-collapse of Pop III stars, they may have already grown into SMBHs and be rare;  in this scenario, IMBHs are
the potential seeds of SMBHs~\cite{Vandeven1991,greene2020intermediate}. IMBHs are excellent candidates to explain the observed Ultra Luminous X-ray (ULX) sources, even though many ULXs have been identified with low-mass X-ray binaries \citep{2014Natur.514..202B} and high-mass X-ray binaries \citep{2014Natur.514..198M,2013Natur.503..500L}. There seems to be, however, a good candidate for an IMBH in a ULX where using high-frequency QPOs (with 3:2 ratio), \citet{2014Natur.513...74P} used inverse-scaling of sBH as well as a relativistic precession model to determine a mass of $M_{BH} \simeq 400 M_\odot$. 

IMBHs impact on several fields of astrophysics and are likely to grow as a focus of research attention \cite{Vandeven1991}. A standard scenario is that these BHs result from accretion onto BH seeds or PBHs. The growth of PBHs could explain the masses of IMBH since these can grow up to $10^2 - 10^5 M_\odot$ during the radiation dominated era \cite{Lora2013b}.  Moreover, in the early time of BH formation, dark matter makes BHs grow and facilitates IMBHs  formation \cite{Lora2013c}. Ref. \cite{volonteri2005dynamical} studied the dynamical evolution of IMBHs in the nearby Universe, in the context of popular hierarchical structure formation theories. This computational analysis  estimated the number of BHs, their mass distribution and position within a galaxy. Additionally, Ref. \cite{gair2011exploring} discussed the capability of a third-generation ground-based detector, such as the Einstein Telescope (ET), to enhance the astrophysical knowledge through detections of gravitational waves emitted by IMBH binary systems. 

An important parameter associated with astrophysical BHs is their angular momentum $J_{BH}$, which is usually described by the dimensionless spin parameter $a_{\star} \equiv a/M_{BH} = c J_{BH} /   G M_{BH}^2$, where $M_{BH}$ corresponds to the BH mass. The spin is related to the way BHs grow and  explains the relativistic jets that emerge from the BH. To model the BH spin up due to plasma accretion,  the Bardeen model \cite{Bardeen:1970} is often used in  the literature, which assumes that the accretion follows from a thin cold disk rotating in the equatorial plane (constant orientation) and that its inner edge always coincides with the ISCO (ballistic or geodesic model). Bardeen found that this process achieves the maximal rotation $a_{\star} =  1$; however, this is an artifact of the model. As Thorne \cite{Thorne1974} pointed out, the spin up all the way to an extreme Kerr BH, using a finite amount of rest mass, violates the third law of BH thermodynamics. Thus, some processes ignored in Bardeen’s simplistic model should keep $a/M_{BH} < 1$. To tackle this shortcoming, Thorne included in the accretion model the absorption of radiation by the BH, observing that the absorption cross-section is greater for photons moving in the opposite direction to the BH's angular momentum than for “co-rotating” photons. The captured photons will lower the efficiency of the BH spin up and thus prevent the violation of the third law; this effect prevents spin-up beyond a limiting value of $a/M_{BH} = 0.998 $. Still, one  must be aware that even the improved Thorne model is based on geodesic motion and on a thin accretion disk, not taking into account  more realistic physics, like the kinetic pressure tensor, magnetic fields, viscosity, density and velocity gradients of the environment, radiation terms, among others. As an  illustration  of the impact of these extra factors,  it was recently shown that the power extracted via fast magnetic re-connection can induce a significant decrease of the rotational energy of the BH, reducing the dimensionless spin from $a_{\star} = 0.999$ to $a_{\star} = 0.99$ \cite{Comisso2021}.

This article aims to explore the initial masses and spins as well as the luminosity of the progenitors of the gravitational wave event GW190521, through a relativistic mass accretion formula, which includes the effects of the density gradients associated with the environment where the massive binary system is immersed. Concretely, we have computed the masses, the bremsstrahlung bolometric luminosity and  the spin as a function of the red-shift, for the individual components of the GW190521 binary system estimated by the LVC collaboration. Our results exhibit, depending on the initial Mach number and density gradient parameter, a wide range of values for the initial masses, which consequently admit a variety of interpretations, from seed BHs formed from Pop III stars collapse, to PBHs. We also perform a comparison between the Newtonian and relativistic regimes, showing that relevant corrections for the initial masses are introduced by the relativistic treatment. Finally, we provided an evolution equation to compute the BH spin, which is written as function of the mass accretion rate and the angular momentum rate. This formula was calculated from the definition of the dimensionless BH spin and using semi-analytical accretion formulas from General Relativity hydrodynamics simulations. It is worth mentioning that we explore self-consistent solutions with very high mass accretion rates, where the accretion is not suppressed by any radiative mechanism like  Compton heating. Such accretion regime is  referred to as the \textit{super-Eddington} regime. This high accretion process has been studied in spherically symmetric accretion flows on to massive BHs embedded in dense metal-poor clouds, performing one-dimensional radiation hydro-dynamical simulations \cite{Kohei2016,sakurai2016hyper}. The accretion rate reached in \cite{Kohei2016,sakurai2016hyper}, in which radiation mechanisms were included, are quite similar to the ones obtained in this work. General relativistic radiation-hydrodynamic simulations of accretion flows onto BHs show that the radiation term can suppress the accretion rate up to two order of magnitude
\cite{Zanotti2011}. Still, most of the aforementioned simulations show a super-Eddington accretion flows.

This article is organized as follows: In section \ref{sec:Acc}, we present the relativistic accretion model as well as the general relativistic equations used in our calculations. In section \ref{sec:Results}, we present the results. Finally, in section \ref{sec:Conclusions}, we present our main conclusions and discussions.

\section{General relativistic accretion model}
\label{sec:Acc}
\paragraph{Plasma accretion rates.} We consider that most of the gas is concentrated in a thin disc and that most of the accretion occurs in the equatorial plane. Under this assumption, the accretion rates have been computed by performing systematic relativistic hydrodynamics simulations of BHL accretion \cite{Lora2015219,Cruz2020b}. We have used the relativistic BHL accretion model for a realistic astrophysical scenario, to describe, say, the common envelope phase of a binary. Such a scenario could occur after the collapse of a neutron star to a black hole in a binary system initially containing red supergiants. The effects of the common envelope are introduced by the dimensionless density gradient parameter $\epsilon_{\rho}$, which runs from $\epsilon_{\rho}=0$ for cosmological or intergalactic gas before the common envelope phase, to $\epsilon \geq 1$ for short orbits when the binary is near the merger \cite{MacLeod2015, Cruz2020b}. 

Our accretion model differs from the one presented by Thorne \cite{Thorne1974} in the fact that it is not geodesic and considers a pure baryonic gas accretion with density and velocity gradients, where such a gas is described by an ideal gas equation of state $p=\rho \epsilon(\Gamma-1)$, which describes the pressure as function of rest mass density $\rho$, specific internal energy $\epsilon$ and the adiabatic index $\Gamma$, neglecting the radiation accretion and magnetic fields \footnote{We are only considering red-shifts up to $z=100$ and the cosmological radiation era occurs for $z\gtrsim 130$.}. These simulations yield a semi-analytical model for the evolution of the rest-mass $M_0$,  specific angular momentum (here denoted $P^\phi$) of the gas and bremsstrahlung luminosity $L_{_{\rm BR}}$ \cite{Gammie03},\footnote{The semi-analytical formulas were fitted by using the following definitions at steady state: Rest-mass accretion rate $\dot{M}_0 = \int \sqrt{-g}\rho u^{i}ds_{i}$, the specific angular momentum rate $\dot{P}^{\phi}=\int \sqrt{-g}T^{i}_{\phi}ds_{i}$, and bremsstrahlung luminosity $L_{\rm BR}=3\times 10^{78}\int (T^{1/2}\rho^{2}\sqrt{\gamma})dV (M_{\odot}/M_{\rm BH})erg/s$, where $T$ is the temperature of the gas, $u^i$ are the spatial component of the four-velocity of the fluid, $T^\mu_\nu$ is the energy-momentum tensor (see all details in \citep{Cruz2020b}), $\sqrt{-g}$ and $\sqrt{\gamma}$ are determinants of the four-metric and spacial metric, respectively.} as a function of the plasma initial density $\rho_{\infty}$, velocity $v_{\infty}$, sound speed $c^{2}_{s, \infty}$, adiabatic index $\Gamma$, dimensionless density gradient parameter $\epsilon_{\rho}$, and mass of the central BH, $M_{\rm BH}$ as follow

\begin{eqnarray}
  \log\left(\frac{\dot{M}_0}{\dot{M}_{0,\rm ref}}\right)&=& \mu_{1} +
  {\mu_{2}}/({1+\mu_{3}\epsilon_{\rho} + \mu_{4}\epsilon^{2}_{\rho}})\,, 
  \qquad \qquad \dot{M}_{0,\rm ref}= 4 \pi \lambda \rho_{\infty} M_{BH}^{1/2} r^{3/2}_{\rm acc}
  \label{eq:fitmrate}\,, \\
  \log \left(\frac{\dot{P}^{\phi}}{\dot{P}_{\rm ref}}\right) &=&
  \wp_{1} +\wp_{2}\epsilon_{\rho} + \wp_{3}\epsilon^{2}_{\rho}\,, 
  \qquad \qquad \qquad \qquad  \dot{P}_{\rm ref}= \dot{M}_{0,\rm ref}\frac{v_{\infty}}{\sqrt{1-v^{2}_{\infty}/c^{2} }}\,, \\
    \log \left(\frac{L_{_{\rm BR}}}{L_{\rm Edd}}\right) &=& \ell_{1}+\ell_{2}\epsilon_{\rho} + \ell_{3}\epsilon^{2}_{\rho}\,.
\end{eqnarray}

Here, the normalization of accretion rates $\rm \dot{M}_{\rm 0, ref}$ and $\rm \dot{P}_{\rm ref}$ are the relativistic spherical mass and angular momentum accretion rates, respectively \cite{Novikov:1973,Petrich89}. It is worth mentioning that the above expression are computed at the BH's event horizon. In the above relations we introduced the accretion radius 
$r_{\rm acc} := GM_{BH} / \left(c^{2}_{s, \infty} + v^{2}_{\infty}\right)$ and the parameter $\lambda := 0.71\ (\Gamma=4/3)=0.25\ (\Gamma=5/3)$; in our model we use the adiabatic index $\Gamma=5/3$.\footnote{$\lambda$ comes from the mass accretion rate of relativistic spherical accretion (analytical solution). The adiabatic index corresponds to an ionized hydrogen plasma with non-relativistic protons and non-relativistic electrons.} The luminosity is normalised by the Eddington luminosity $\rm L_{\rm Edd} = 1.26\times 10^{38}\ (M_{BH}/M_{\odot})\ erg/s$ \cite{Rezzolla_book:2013}. The coefficients in the accretion formulas are: for the mass accretion rate, $\rm \mu_{i}=(0.18,1.24,-1.54,0.72)$; for the angular momentum accretion rate, $\rm \wp_{i}=(0.51, 6.69, 1.16)$; and for the  bremsstrahlung  luminosity, $\rm \ell_{i}=(-3.2837,3.1742,3.327)$. The dimensionless density gradient parameter range considered is $0\leqslant \epsilon_{\rho} \leqslant 1$ \cite{Lora2015219,Cruz2020b}. In the strong gravity regime, i.e., the relativistic regime,  BHL studies reveal super-Eddington accretion rates, even in more realistic models where the self-consistent radiation is considered \cite{Zanotti2011,Roedig2011}; in the same scenarios luminosity can be sub-Eddington. In our accretion model, $\epsilon\geq 0.5$ leads to super-Eddington luminosities. However, in early universe scenarios, i.e., with high red shifts $z\geq 10^{3}$, radiation has an important contribution to the accretion, and the Eddington-accretion limit should be monitored. Then, the black hole size should be constrained to be smaller than the cosmological particle horizon (size of the universe) \cite{Rice2017}. 

The spin, or Kerr parameter, $a_{\star}$,  varies between $-1$ for accretion disks which counter-rotate with respect to the BH, it is $0$ for non spinning BHs, and it is $+1$ for the co-rotating case. Since we are measuring the rest-mass and the angular momentum accretion rates close to the event horizon, we assume that all is accreted by the BH, that is $\dot{M}_{BH} =- \dot{M}_0 $ and $\dot{J}_{BH} =-r_{\rm EH} \dot{P}^{\phi}$. This leads to the following law for the BH spin evolution due to gas accretion \cite{Gammie:2003qi}
\begin{eqnarray}
\dot{a}_{\star} = 
\frac{c}{G}\frac{\dot{J}_{BH}}{M_{BH}^{2}} - 2 a_{\star}  \frac{\dot{M}_{BH}}{M_{BH}},
\label{eq:adot}
\end{eqnarray}
where dot denotes derivative with respect to time. It is worth mentioning that mass and spin accretion rates given in equations (\ref{eq:fitmrate}--\ref{eq:adot}), as well as the event horizon radius, are computed in quasi-stationary states. In the forthcoming calculations of binary evolution, we assume that BH's are growing in a sequence of stationary states, which means that we track the accretion process during this time domain as a sequence of steady processes of accretion during finite time intervals $t \in  [t_i, t_{i+1}]$ such that $t_0 < t_1 < ... <t_{N-1} < t_f$, allowing to update the mass, spin, event horizon radius and accretion rates at every time step, see \cite{Lora2013b}.

\begin{figure}[h!]
\centering 
\includegraphics[width=.5\textwidth,angle=0]{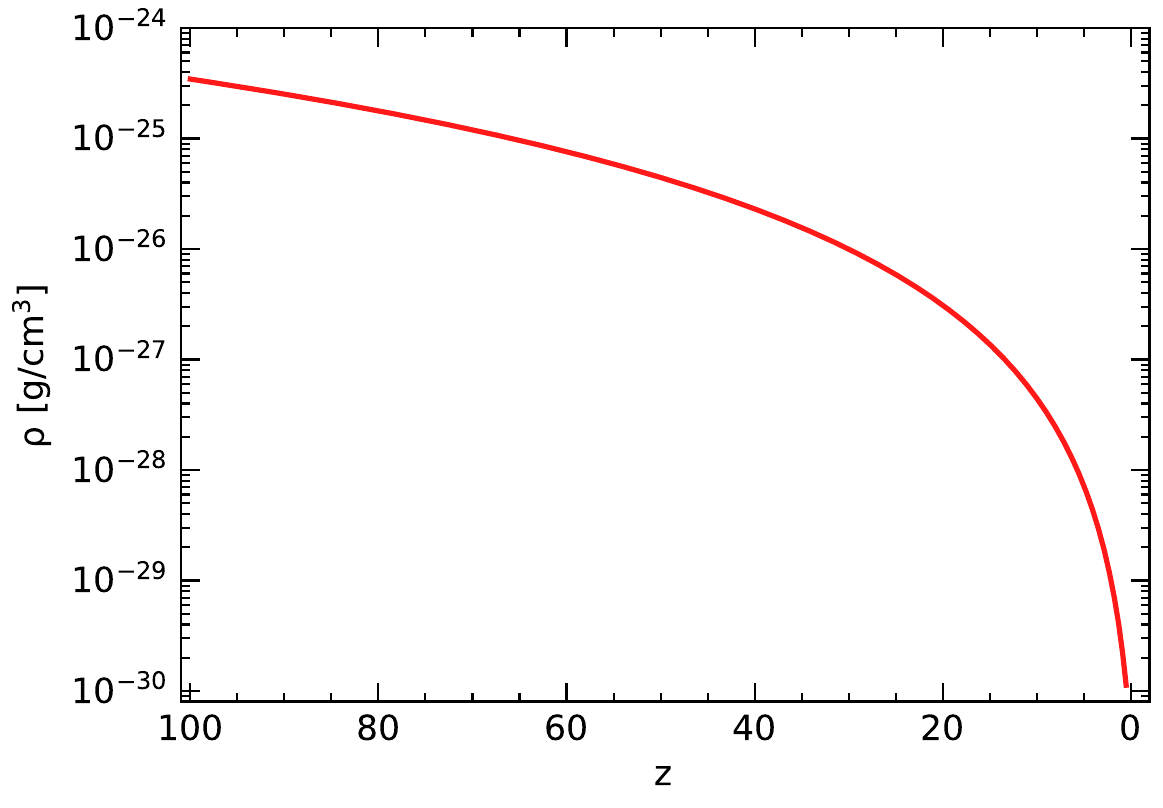}\hspace{-0.05cm}
\includegraphics[width=.495\textwidth,angle=0]{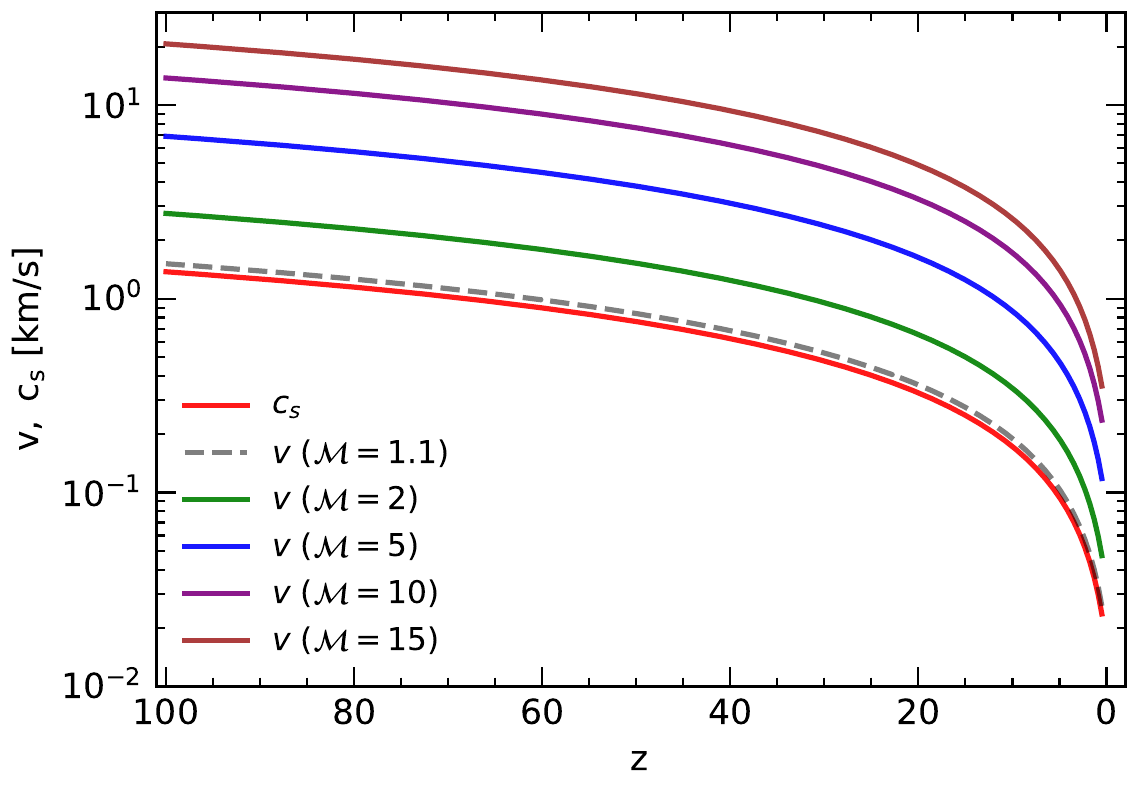}\\
\caption{\label{fig:Den} Rest mass density (left) and sound speed (right) profiles of the cosmic plasma as functions of  the redshift $z$, given by \eqref{eq:den} and \eqref{eq:cs}. The right panel also shows some representative cases for the gas velocity evolution,  assuming constant relativistic Mach numbers ${\cal M}= 1.1,~2,~5,~10$, and $15$. The $x$-range covers only the redshifts relevant for our analysis.}
\end{figure}

\paragraph{Intergalactic density and sound speed evolution.} The evolution of the cosmic plasma density and sound speed after the recombination ($z\sim 10^{3}$  - assuming a pure hydrogen gas) can be defined as a function of the red-shift $z$, in cgs unit, as follow

\begin{eqnarray}
\rho(z)&=& 200 m_{H}  \left(\frac{1+z}{10^{3}}\right)^{3} \frac{g}{cm^{3}}\,,  \label{eq:den}\\
c_{s}(z) &=& 5.7 \sqrt{\frac{1+z}{10^{3}}} \left[1 + \left(\frac{1+z_{\rm dec}}{1+z}\right)^{\beta} \right]^{-\frac{1}{(2\beta)}}\ \frac{km}{s}\,, \label{eq:cs} 
\end{eqnarray}
where $\beta=1.72$, $z_{\rm dec}=130$ is the red-shift when baryonic matter decouples from the radiation fluid and $\rm m_{H}=1.6735575\times 10^{-24} g$ is the hydrogen mass. The sound speed is defined assuming the plasma temperature decreases in an adiabatic way after $z=100$, when the electron temperature decouples from the cosmic microwave background (CMB) \cite{Ricotti2008, DeLuca2020a}. The density defined in \eqref{eq:den} will be used in  \eqref{eq:fitmrate} and redefined as $\rho_{\infty}$; furthermore, for the model where a density gradient is also considered, such density will be representing the density at the position of the BH, $r_{0}$, which corresponds to the center of the gradient direction (the density gradient is applied in the perpendicular direction to the BH motion) $\rho(r)= \rho_{\infty} {\rm exp}[-\epsilon_{\rho}(r-r_{0})/r_{\rm acc}]$, see Ref. \cite{Cruz2020b} for more details. Meanwhile, the sound speed in \eqref{eq:cs} is used to define the velocity of the plasma surrounding the binary BHs through the relativistic Mach number definition $\mathcal{M}:=W v/W_{s}c_{s}$, where $W$ is the Lorentz factor and $W_s$ is the Lorentz factor of the sound speed. The evolution of the plasma density, sound speed, and a few representative velocities as a function of the red-shift $z$ have been plotted in Figure \ref{fig:Den}, spanning a time interval from the early universe, $z=100$, to the LVC estimated red-shift for GW190521, $z=0.82$.

\paragraph{Accretion into the individual BHs in the binary system.} The mass accretion rate for the individual GW190521 progenitor BHs is measured using the global accretion rate of the total mass of the binary system, following the approximation from \cite{DeLuca2020b}

\begin{eqnarray}
\dot{M}_{1}=\frac{\dot{M}_{\rm bin}}{\sqrt{2(1+q)}}\,, \qquad \dot{M}_{2}=\frac{\dot{M}_{\rm bin} \sqrt{q}}{\sqrt{2(1+q)}}\,, \label{eq:binaryacc}
\end{eqnarray}
where $q=M_{2}/M_{1}\leqslant 1$ is the mass ratio and $\dot{M}_{\rm bin}$ is the accretion rate of the binary system, neglecting the eccentricity contribution and assuming that the orbital period of the binary is much smaller than the accretion time scale. For $z\leqslant 100$ the accretion time scale $\tau_{\rm acc}$\footnote{The accretion time scale is defined as $\tau_{\rm acc}=\sigma_{\rm T}/( {\rm 4 \pi m_{p}} \dot{m})=(4.5\times 10^{8} yr)/\dot{m}$ \cite{DeLuca2020b}, where $\dot{m}=\dot{M}/\dot{M}_{\rm Edd}$, and $\dot{M}_{\rm Edd}=1.39 \times 10^{17} (M/M_{\odot}) g/s$ \cite{Rezzolla_book:2013}.} is smaller than the age of the universe, the accretion can therefore play an important role in the mass evolution of the BHs \cite{Ricotti2008}; thus our calculation will starts at $z=100$.

The procedure to compute the mass evolution of the individual BHs in the binary is as follows:
\begin{itemize}
    \item The binary mass accretion rate $\dot{M}_{\rm bin}$ is computed using the analytical formula given in \eqref{eq:fitmrate} together with the density defined in \eqref{eq:den}, the sound speed presented in  \eqref{eq:cs} and the velocity from the relativistic Mach number.
    \item The individual mass accretion rates, specific angular momentum and luminosity are estimated using the relations given in \eqref{eq:binaryacc}.
    \item The individual mass and spin evolution is computed at each time step as follows:
    \begin{eqnarray}
    M_{1} &=& M_{i,1} + \Delta t\ \dot{M}_{1}\,, \quad M_{2} = M_{i,2} + \Delta t\ \dot{M}_{2}\,,\\
       a_{\star,1} &=& a_{\star\ i,1} + \Delta t\ \dot{a}_{\star,1}\,, \quad a_{\star, 2} = a_{\star,i,2} + \Delta t\ \dot{a}_{\star, 2}\,,
    \end{eqnarray}
    where the index $i$ indicates the values at the initial time, and $\Delta t$ is the time step of the universe evolution written in terms of the red-shift and the Hubble constant parameter $\rm H_{0}= 67.9\ (km s^{-1})\ Mpc^{-1}$. 
\end{itemize}
We investigate the binary BH evolution in the red-shift range $\rm z\in [100,0.82]$; as already mentioned, $z=0.82$ corresponds to the GW190521 merger time estimated by the LVC, at  which point the final masses and spins (following the LVC estimates  \cite{Abbott:2020tfl}) are 
\begin{equation}
  \rm M_{1}=85^{+21}_{-14}\ M_\odot \ , \quad  \rm M_{2}=66^{+17}_{-18}\ M_\odot \ , \qquad \rm a_{\star,1}=0.69^{+0.27}_{-0.62} \ , \quad a_{\star,2}=0.73^{+0.24}_{-0.64} \ .
\end{equation}

Our resolution time step is defined by the step in the redshift, $\Delta z= 10^{-4}$. This means that the mass backward integration have been computed over $\sim 10^{6}$ time steps. The parameter space analyzed in this paper consists in $1.6\times 10^{5}$ accretion scenarios varying the relativistic Mach number ${\cal M}\in [1,10]$, covering some astrophysical scenarios as: common envelope, $2 \leq {\cal M} \leq 7$ \cite{Cruz2020b}; Pop Star III scenarios, $ {\cal M} \leq 25$ \cite{Kinugawa2017}; and some cosmological accretion models, $2 \leq {\cal M} \leq 10^{3}$ \cite{Ricotti2008}. The range for the density gradient  $\epsilon_{\rho}\in [0,1]$ in the plasma is motivated by the stellar evolution model.

\section{Results} \label{sec:Results}
\begin{figure}[tbp]
\centering 
\includegraphics[width=.95\textwidth,angle=0]{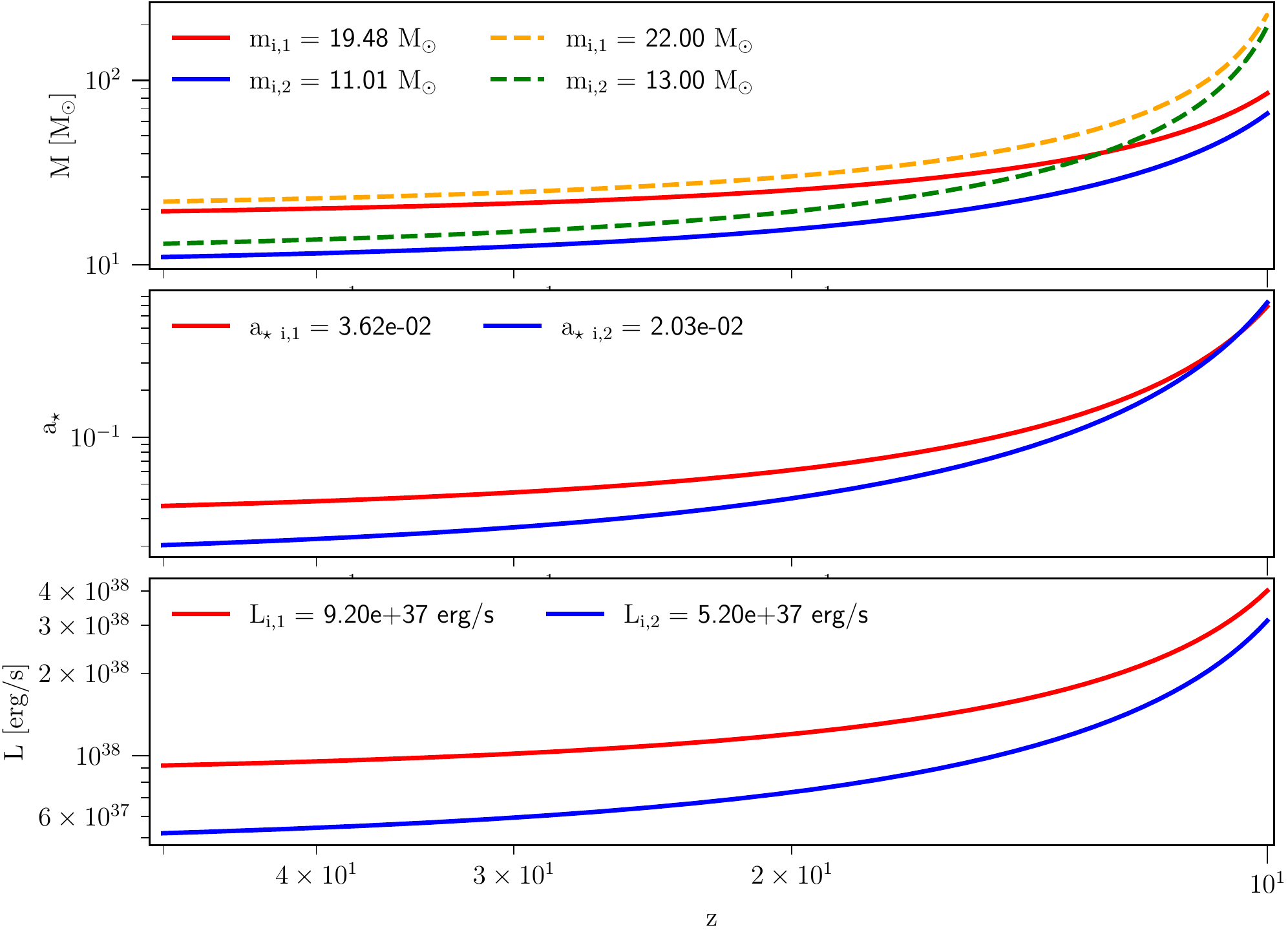}
\caption{\label{fig:Newt} Evolution of the BH mass (top), BH spin (middle) and bremsstrahlung bolometric luminosity (bottom) of the GW190521 progenitors, $vs.$ the red-shift, using the relativistic accretion model. Here, we assume that accretion process takes place between $\rm z=50$ and $\rm z=10$. The blue and red lines shows the evolution that match the LVC estimated parameters at the end of the accretion ($M_{1}=85 M_{\odot}$, $M_{2}=66M_{\odot}$ and $a_{\star, 1}=0.69$, $a_{\star, 2}=0.73$). The orange and green lines correspond to evolutions starting  with the progenitors masses $m_{i,1}=22M_{\odot}$ and $m_{i,2}=13M_{\odot}$ that, according to  the Newtonian accretion model \cite{DeLuca2020}, match the LVC estimated parameters at the end of the accretion; such initial data clearly overshoots the GW190521 parameters, under a relativistic evolution. For this example we took the interstellar gas moving with twice the sound speed (Mach two) with respect to the BH frame.}
\end{figure}
\begin{figure}[h!]
\centering 
\includegraphics[width=.95\textwidth,angle=0]{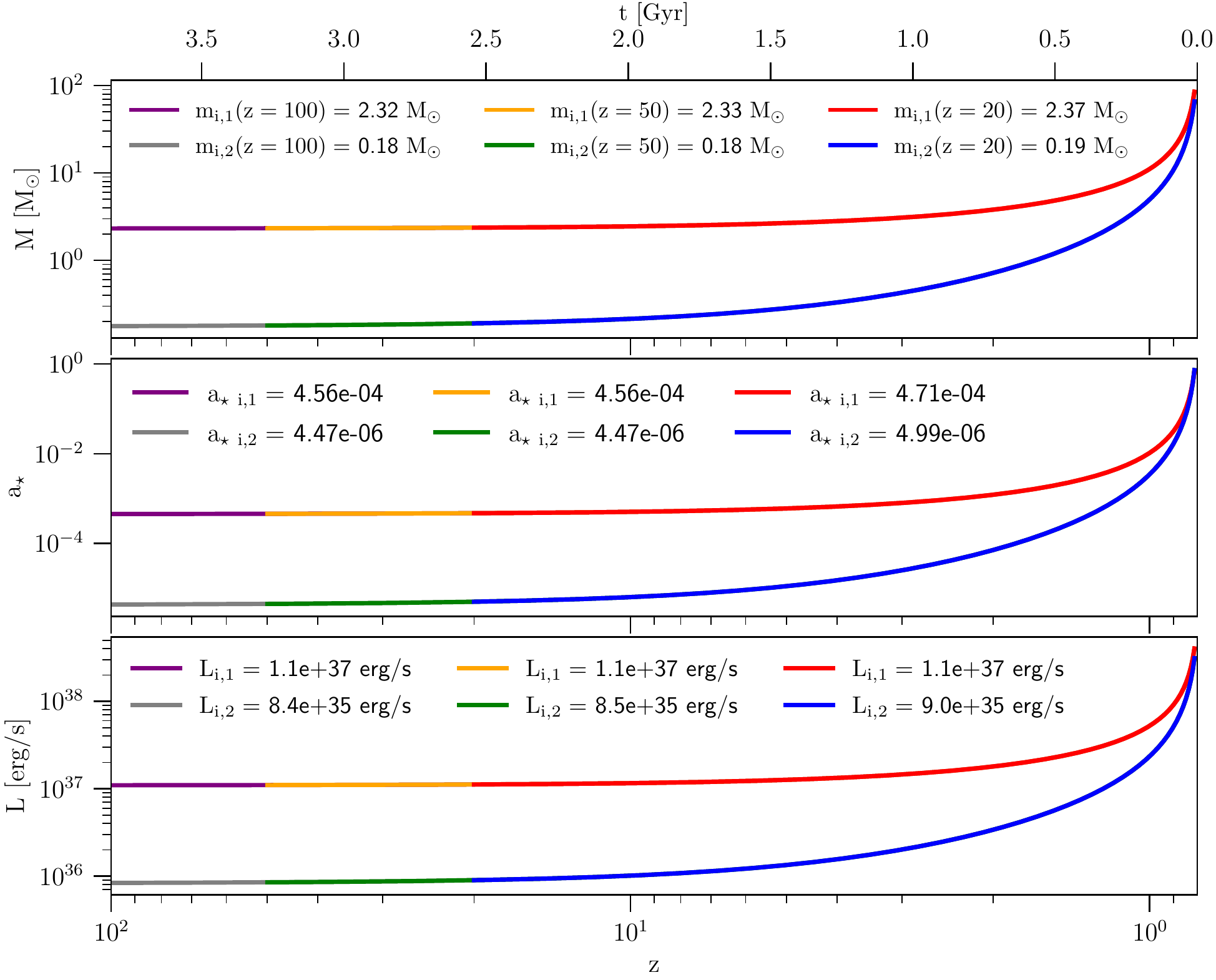}
\caption{\label{fig:twoz} Evolution of the BH mass (top), spin of the BH (middle) and bremsstrahlung bolometric luminosity (bottom) of the GW190521 progenitors, $vs.$ the red-shift, using the relativistic accretion model. Here the evolutions assume the final data given by the LVC at $z=0.82$ for GW190521. The backwards evolution goes back to $z=20$ (red and blue lines),  $z=50$ (orange and green lines) or $z=100$ (purple and gray lines). The results shows that the BHs gain less than $2\%$ of their masses from red-shifts $100-20$.}
\end{figure}

\paragraph{Comparison between the relativistic and Newtonian models.} In Figure \ref{fig:Newt} we show the evolution of: the BH masses, in solar masses units (top panel), BH spins, normalized to the initial spins (middle panel) and bremsstrahlung bolometric luminosity (bottom panel) of the individual BHs in the binary. To perform a comparison between the previous results using the Newtonian model and our relativistic accretion model we proceeded as follows. First, we have performed a forward integration (with our relativistic model) using the initial BH masses ($m_{i,1}=22M_{\odot}$ and $m_{i,2}=13M_{\odot}$) previously obtained using the Newtonian model~\cite{DeLuca2020}, that,  according   to the latter, lead to final masses at $z=10$ coinciding with the LVC parameter estimation. For a fair comparison, we have only considered red-shifts in the range $z\in [50,10]$, gas Mach number ${\cal M}=2$ and no density gradients ($\epsilon_{\rho}=0$). The green and orange lines (top panel) show such evolutions; the final masses are over 200\% larger than those estimated for GW190521. Second,  we have performed a backward integration using the final LVC parameters; the blue and red lines show such evolutions, from where we found the initial masses $m_{i,1}=19.5M_{\odot}$ and $m_{i,2}=11M_{\odot}$; these are $\sim$ 13\% smaller than the Newtonian estimations. Both models produce low luminosity emissions, less than $10^{39}\ erg/s$.  
\paragraph{Small $vs.$ large red-shift accretion.} Figure \ref{fig:twoz} shows evolutions for three different scenarios for the origin of the  GW190521 progenitors, all obtained by backwards integration  from the  LVC parameters at red-shift $z=0.82$, and assuming that the gas is moving supersonic velocity,  ${\cal M}=2$. The first scenario traces the progenitors back to the early universe, at $z=100$, obtaining initial masses $m_{i,1}=2.32M_{\odot}$ and $m_{i,2}=0.18M_{\odot}$ (purple and grey solid lines); the second scenario traces the progenitors only back to $z=50$, still obtaining similar initial masses (to the first scenario), $m_{i,1}=2.33M_{\odot}$ and $m_{i,2}=0.18M_{\odot}$ (orange and green lines); the third scenario only evolves back to $z=20$, but still  obtains similar masses,  $m_{i,1}=2.37M_{\odot}$ and $m_{i,2}=0.19M_{\odot}$ red and  blue lines). 
Thus, the BHs gain most of their masses in the final stage of the evolution, for red-shifts $z <10 $, where an exponential increment of their masses occurs. Mass (top), spin (middle) and luminosity (bottom) show the same behavior. The putative  different red-shift origin of the progenitors leads to only to differences of $\sim$ 2\% in the initial masses and spins for formation at $z=100,\ 50$ or $20$ red-shifts.
This suggests three different possible origins for the GW190521 progenitors: as primordial BHs at red-shift $z=100$, from the collapse of dense clumps at red-shift $z=50$, or as core-collapse of Pop III stars at red-shift $z=20$.

\begin{figure}[h!]
\centering 
\includegraphics[width=0.9\textwidth,angle=0]{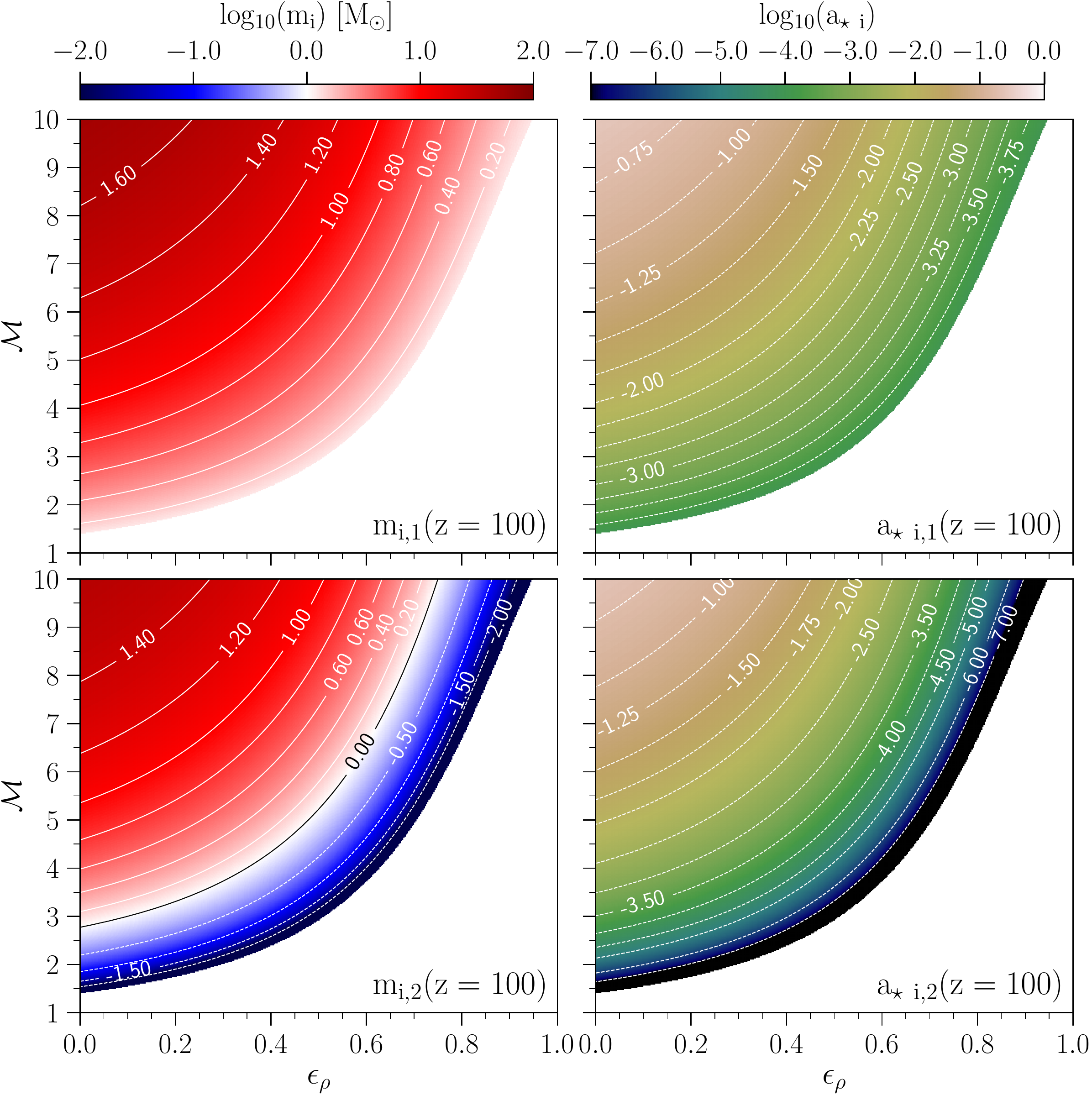}
\caption{\label{fig:2Dsol} GW190521 OPs masses and spins at $\rm z=100$ in logarithmic scale as function of relativistic Mach number $\mathcal{M}$  and density gradient in the interstellar medium $\epsilon_{\rho}$. All points are obtained integrating  back from the  LVC estimated final masses at red-shift $\rm z=0.82$~\cite{Abbott:2020tfl} ($\rm M_{1}=85M_{\odot}$ and $\rm M_{2}=66M_{\odot}$ and $a_{\star, 1}=0.69$, $a_{\star, 2}=0.73$). The plots shows $1.6\times 10^{5}$ solutions varying the interstellar gas velocity and density gradients.}
\end{figure}

\paragraph{Initial BHs mass and spin dependence on the gas velocity and density gradients.} The previous analysis established the insensitivity of the OPs parameters to the cosmological epoch of  formation, as long  as such formation occurs for $z\gtrsim 10$. Let  us now consider the dependence on the details  of the accretion. In Figure \ref{fig:2Dsol}, we show initial masses in solar masses $\rm m_{i,1}$ (top left) and $\rm m_{i,2}$ (bottom left) in logarithmic scale integrating back from the LVC estimated parameters at $z=0.82$   back  to  red-shift $\rm z=100$. The right panels exhibit the corresponding spins $a_{\star i,1}$ and $a_{\star i,2}$.
 The plots were obtained from  $\rm 1.6\times 10^{5}$ accretion models, varying the relativistic Mach number of the gas, considering supersonic relative motion with Mach numbers in the range of ${\cal M}\in [1,10]$, and density gradient in the plasma $\epsilon_{\rho}\in [0,1]$.

The backward integration of the relativistic accretion equations reveals that some models are self-constrained. Observe the white regions in the right bottom part of the plots; this excluded region is delimited by small mass BHs, from  a few to $10^{-3}$ solar masses (greater than the maximum mass for which Hawking evaporation becomes relevant). Such BHs are only allowed if formed in the early universe as PBHs. On the other hand, for low-density gradients $\epsilon_{\rho} < 0.4$ and high Mach number ${\cal M}>8$ we have a family of high mass BHs; in such models the progenitors keep an almost constant mass from their formation up to the merger at $z=0.82$, accreting only a very small fraction of their mass. {The largest initial masses found are $m_{i,1}\sim 71 M_{\odot}$ and $m_{i,2}\sim 54 M_{\odot}$ with spins $a_{i,\star}\sim 0.48$ and $a_{i,\star}\sim 0.49$, respectively, around $80\%$ of the final masses and around $68\%$ of the final spins.} 

In Table \ref{tab:massandspins} we display nine representative models, varying the density gradient and Mach number for each formation scenario. Our relativistic accretion model also allows sBHs as the  OPs, for some range of density gradients and Mach numbers. Such binary systems can originate from the collapse of dense clumps and core-collapse of hyper-massive stars, gaining their masses through accretion and reaching the observed masses. The bolometric luminosity generated by the accretion process is of the order of $\rm 10^{39}\ erg/s$ in all cases, at the merger. This low  value is consistent with the LVC conclusion that no electromagnetic emission was detected. 

\begin{table}[h!]
\centering
\caption{Representative values for the mass and spin of the  GW190521  OPs at $z=100,\ 50,$ and $20$, by assuming that the final quantities match the  LVC values at $z=0.82$. As illustrations, we show the models for density gradients of the interstellar medium $\epsilon_{\rho}=0, ~0.2$ and $0.4$, and three relativistic Mach numbers ${\cal M}=2,\ 6,\ 10$, and $15$ (the minimum allowed for $\epsilon_{\rho}=0.4$ is slightly higher).}
  \label{tab:massandspins}
\begin{adjustbox}{max width=1.01\textwidth}
\begin{tabular}{cc|ccc|ccc|ccc|ccc} 
 \hline
 $\epsilon_{\rho}$&${\cal M}$&       &$\rm m_{i,1}[M_{\odot}]$&    &       &$\rm m_{i,2}[M_{\odot}]$&      & &$\rm a_{\star\ i,1}$&    &       &$\rm a_{\star\ i,2}$ \\
 \hline \hline
    &   & $z=100$&  $z=50$& $z=20$ & $z=100$&  $z=50$ & $z=20$ & $z=100$&  $z=50$& $z=20$ & $z=100$&  $z=50$ & $z=20$ \\
 \hline 
    & 2 & 02.32& 02.33& 02.37& 00.18  & 00.19  & 00.20 & 4.52E-4& 4.56E-4& 4.71E-4& 4.33E-6 & 4.47E-6 & 4.99E-6\\
0.0 & 6 & 22.88& 22.97& 23.29& 13.60  & 13.67  & 13.92 & 4.98E-2& 5.02E-2& 5.16E-2& 3.08E-2& 3.12E-2& 3.23E-2\\
    &10 & 51.76& 51.87& 52.27& 37.19  & 37.29  & 37.63 & 2.56E-1& 2.57E-1& 2.61E-1& 2.32E-1& 2.33E-1& 2.37E-1\\
    &15 & 71.11& 71.18& 71.41& 53.84  & 53.89  & 54.10 & 4.83E-1& 4.87E-1& 4.84E-1& 4.86E-1& 4.87E-1& 4.90E-1\\
 \hline    
    &2  & 01.60& 01.61& 01.64& 3.00E-2& 3.00E-2& 3.00E-2& 1.97E-4& 1.99E-4& 2.05E-4& 8.24E-8& 8.70E-8& 1.08E-7\\
0.2 &6  & 16.31& 16.37& 16.62& 08.66  & 08.71  & 08.89 & 2.52E-2& 2.54E-2& 2.62E-2& 1.24E-2& 1.26E-2& 1.31E-2\\
    &10 & 42.05& 42.17& 42.59& 29.04  & 29.14  & 29.49 & 1.69E-1& 1.70E-1& 1.73E-1& 1.41E-1& 1.42E-1& 1.46E-1 \\
    &15 & 64.72& 64.81& 65.12& 48.30  & 48.37  & 48.64 & 4.00E-1& 4.01E-1& 4.05E-1& 3.91E-1& 3.92E-1& 3.96E-1\\
 \hline
    &2.4& 01.25& 01.26& 01.28& 5E-4   & 6E-4   & 1.2E-3& 1.11E-4& 1.12E-4& 1.15E-4& 2.20E-11&3.22E-11&1.12E-10\\
0.4 &6  & 09.16& 09.20& 09.35& 03.73  & 03.76  & 03.85 & 7.88E-3& 7.95E-3& 8.21E-3& 2.27E-3& 2.30E-3& 2.42E-3\\
    &10 & 27.37& 27.46& 27.83& 17.11  & 17.19  & 17.47 & 7.14E-2& 7.19E-2& 7.38E-2& 4.89E-2& 4.93E-2& 5.10E-2 \\
    &15 & 51.22& 51.33& 51.73& 36.74  & 36.83  & 37.17 & 2.50E-1& 2.52E-1& 2.56E-1& 2.26E-1& 2.27E-1& 2.32E-1\\

 \hline
 \end{tabular}
 \end{adjustbox}
\end{table}

\begin{figure}[h!]
\centering  
\includegraphics[width=1.0\textwidth]{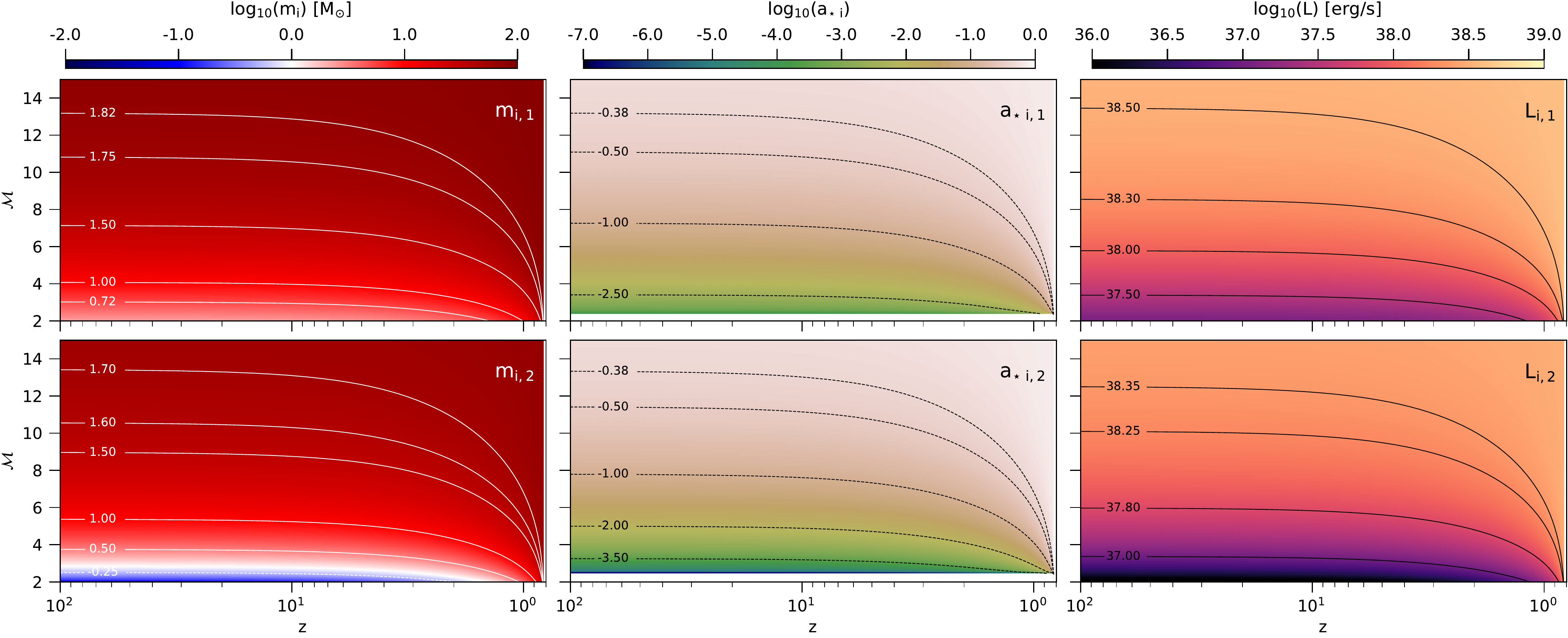}
\caption{\label{fig:2Ddiagrams} Evolution of the BH mass (left), BH spin (middle) and bremsstrahlung luminosity (right) in logarithmic scale for the two component of the binary BH system. Each panel shows the evolution of this physical quantities in the red-shift range $\rm z\in[100,0.82]$ varying the relativistic Mach number, note that initial masses of progenitors at $z=100$ are also increasing as we increase the Mach number (representative values presented in Table \ref{tab:massandspins}). Contour lines represent constant values for each quantity in logarithmic scale. Top panels corresponds to first BH component and bottom panels to the second binary component quantities. For all models we are assuming zero density gradients.}
\end{figure}

{Figure \ref{fig:2Ddiagrams} shows the OPs masses (left panels), OPs spins (middle panels) and luminosity, for homogeneous interstellar gas i. e. $\epsilon_{\rho}=0$. Top and bottom corresponds to each component of the binary. The secular evolution shows a tiny increment at early stages ($z<5$) for low Mach numbers, ${\cal M} < 5$, and an exponentially growth only in the last stage of the evolution. For high Mach numbers ${\cal M} > 5$, we observe a smooth growth in all quantities. By backward integration starting with LVC observations, we found that the progenitors in these cases only accreted $\sim 20\%$ of its masses and increased only  $\sim 32\%$ of its original spins. Overall, we found that the luminosity due to heated plasma in the accretion process is increasing around three orders of magnitude.}

\section{Discussion and Conclusions}
\label{sec:Conclusions}

Motivated by the recently reported GW190521 gravitational wave transient, and the fact that the estimated masses of the two BH progenitors for this event fall into the so-called pair instability supernova gap, in this paper we have studied the role of accretion, in  a relativistic treatment, in the evolution  of  these progenitors. Thus, we have explored the possible initial masses  and spins for the progenitors, as well as the luminosity of the IMBH detected from the gravitational wave event GW190521, by using a general relativistic mass and spin accretion formula. These formulas allow to include the effects of the density gradients associated with the environment where the massive binary system is immersed. The properties of the accretion flow and spacetime --horizon penetrating ingoing coordinates -- also allows stable accretion rates above the Eddington limit \cite{Cruz2012}. Furthermore, models with $\epsilon_{\rho}\leq 0.5$ and self-consistent radiation evolution exhibit sub-Eddington luminosities, while mass accretion rates are always super-Eddington \cite{Zanotti2011,Cruz2020b}. Such high accretion rates (super or hyper Eddington) $\dot{M} \sim 5 \times 10^{4} \dot{M}_{\rm Edd}$ can occur in scenarios where the intermediate BH is surrounded by dense metal-poor clouds, also possible when the electron scattering produced by trapped photon close to the BH becomes important \cite{Kohei2016}. The imprints on the dynamic gas-rich environments in the stellar origin of the progenitor could be detectable in next-generation x-ray and radio observations, by multiwavelength LISA detector \cite{Caputo2020}. Our semi-analytical formulas can be improved by performing realistic general-relativistic radiation-hydrodynamic simulations at radiation-pressure-dominated regime, where accretion rates can be reduced two orders of magnitude smaller than in the purely hydrodynamic case \cite{Zanotti2011}. However, we expect that the radiation field will reduce, but not suppress, the super-Eddington accretion, and we anticipate super-Eddington accretion of the order of $\dot{M}  \sim 10^{2}\dot{M}_{\rm Edd}$.

Concretely, we have computed the masses, the spins and the bremsstrahlung bolometric luminosity as a function of the red-shift of the epoch of formation of  the progenitor BHs. The results allow a wide range of values for the initial masses and spins, depending on the initial Mach number and density gradient parameter, which can thus originated from different astrophysical scenarios at different red-shifts. We have carried out a comparison between the initial masses of the GW190521 binary system, predicted using the Newtonian and relativistic expressions of the mass accretion rates. For this comparison, we assume that the BHs formed  at $\rm z=50$ and merged at $\rm z=10$ with final masses $\rm M_{1}=85 M_{\odot}$ and $\rm M_{2}=66M_{\odot}$. In the Newtonian case, the initial masses found for the individual progenitors were $\rm m_{i,1}=22M_{\odot}$ and $\rm m_{i,2}=13M_{\odot}$, while using the relativistic formula the initial masses were $\rm m_{i,1}=19.5M_{\odot}$ and $\rm m_{i,2}=11M_{\odot}$, which are $\rm 13\%$ smaller than in the Newtonian case. Moreover, if we use the relativistic accretion formulas to evolve the initial masses previously estimated with the Newtonian treatment, $\rm m_{i,1}=22M_{\odot}$ and $\rm m_{i,2}=13M_{\odot}$ \cite{DeLuca2020a}, the final masses of the system are $\rm 200\%$ larger than the LVC estimated ones. These calculations mean that important corrections are introduced by considering the relativistic formula \cite{Cruz2020b}, as  expected, since these are strong gravity systems. It is worth mentioning that in order to perform this comparison we did not include density gradients ($\epsilon_{\rho}=0$), since Newtonian mass accretion rate does not include this kind of information about the environment wherein the massive binary system is immersed. 

The main features of our results have been quantified by analyzing the initial masses, spins and the bremsstrahlung bolometric luminosity of the GW190521 OPs, as functions of the relativistic Mach number and  of the density gradients of the environment were the system is immersed. Here we assume that the observed masses $\rm M_{1}=85 M_{\odot}$ and $\rm M_{2}=66M_{\odot}$ with spins $a_{\star,1}=0.69$ and  $a_{\star,2}=0.73$ were measured at $\rm z=0.82$ consistently with the LVC parameter  estimation for GW190521. We have found from relativistic accretion calculations that the BHs detected from GW190521 could originate at considerably different red-shifts,  say, $\rm z=20, \ 50$, and $\rm 100$, with only small variations in their mass and their spin, of $\rm \sim 2\%$. Concerning to the bolometric luminosity, it showed an increment from $\rm z=100$ up to $\rm z=0.82$, of around three orders of magnitudes  (see Figure \ref{fig:2Ddiagrams}). Still, the maximum bolometric luminosity generated by the relativistic accretion process is of the order of $\rm 10^{39}$erg/s, which correspond to low energetic emission. This result is consistent with the LVC observation of   GW190521 \cite{Abbott:2020tfl}, where no electromagnetic emission was detected.

We also found self-constrained models, corresponding to the white regions in Figure \ref{fig:2Dsol}, where the backwards integration gives forbidden solutions, i.e., negative masses. Such constrained region is correlated with the density gradient: increasing the density gradient the constrained regions also increases, allowing only the models where the Mach numbers are high. This region is bounded by OPs with small masses and spins, of $\rm m_{i,1}\sim 10^{-1} M\odot$ and $\rm m_{i,2}\sim 10^{-4} M\odot$, at $\rm z=100, \ 50$, and  $\rm2 0$. Thus, PBHs that grow through accretion could  be the origin of the GW190521 progenitors. The growth of PBHs corresponds to a possible scenario to explain the masses of IMBH since these can grow up to $\rm 10^2 - 10^5 M_\odot$ during the radiation dominated era \cite{Lora2013b}. On the other hand, for low values of the density gradient parameter, $\epsilon_{\rho}<0.4$ and high values of the Mach number, ${\cal M}>8$, we found a family of massive initial BH seeds, with masses of $\rm m_{i,1}=71M_{\odot}$ and $m\rm _{i,2}=54M_{\odot}$ with dimensionless spins $a_{\star}=0.48$ and $a_{\star}=0.49$. In this  case, the accretion rates are small due to the high Mach numbers. The early formation of BHs, with these initial masses, can be associated with the collapse of Pop III stars that occurs at redshift $\rm  15 < z < 30$ \cite{coleman2004intermediate, greene2020intermediate}. Several solutions found correspond to the slowly rotating sBHs seeds, $\rm m_{i} \sim 3-40 M\odot$, which could form at early stages by direct collapse, and grow in mass due to  accretion until reaching  the final values at the merger compatible  with the LVC parameter estimates.

In summary, the relativistic accretion model permits qualitatively different OPs:  non-rotating primordial BHs, with initial masses $ M_{i} \lesssim 1 M\odot$ originated at high red-shifts $z \lesssim 100$; slowly rotating ($a_{\star} < 0.5$) stellar mass BHs with  $\rm 3M\odot \lesssim M_{i} \lesssim 40 M\odot$ originated from stellar collapse; and moderately rotating ($a_{\star} \sim 0.5$) intermediate-mass BHs $\rm 40M\odot \lesssim M_{i} \lesssim 70 M\odot$ from the collapse of Pop III stars.

\acknowledgments
We thank  Paolo Pani for an important comment concerning the spin evolutions on the first version of this paper. ACO gratefully acknowledges support from the COST Action CA16214 ``PHAROS", the  LOEWE-Program in HIC for FAIR, and the  EU Horizon 2020 Research ERC Synergy Grant ``Black-HoleCam: Imaging the Event Horizon of Black Holes" (grant No. 610058).  F.D.L-C was supported by Vicerrectoría de Investigación y Extensión - Universidad Industrial de Santander, under Grant No. 2493. This  work  is also supported  by  the Center  for  Research  and  Development  in  Mathematics  and  Applications  (CIDMA)  through  the Portuguese Foundation for Science and Technology (FCT - Funda\c c\~ao para a Ci\^encia e a Tecnologia), references UIDB/04106/2020 and UIDP/04106/2020.  We acknowledge support  from  the  projects  PTDC/FIS-OUT/28407/2017,  CERN/FIS-PAR/0027/2019 and PTDC/FIS-AST/3041/2020.   This work has further been supported by the European Union’s Horizon 2020 research and innovation (RISE) programme H2020-MSCA-RISE-2017 Grant No. FunFiCO-777740.
The authors would like to acknowledge networking support by the COST Action CA16104.

\bibliographystyle{apsrev}           
%\bibliography{aeireferences}

\end{document}